# On the role of technology in human-dog relationships: a future of nightmares or dreams?

Dirk van der Linden, Brittany I. Davidson, Orit Hirsch-Matsioulas, and Anna Zamansky

*Abstract*—Digital technologies that help people take care of their dogs are becoming more widespread. Yet, little research explores what the role of technology in the human-dog relationship should be. We conducted a qualitative study incorporating quantitative and thematic analysis of 155 UK dog owners reflecting on their daily routines and technology's role in it, disentangling the what-where-why of interspecies routines and activities, technological desires, and rationales for technological support across common human-dog activities. We found that increasingly entangled daily routines lead to close multi-species households where dog owners conceptualize technology as having a role to support them in giving care to their dogs. When confronted with the role of technology across various activities, only chores like cleaning up after their dogs lead to largely positive considerations, while activities that benefit themselves like walking together lead to largely negative considerations. For other activities, whether playing, training, or feeding, attitudes remain diverse. In general, across all activities both a nightmare scenario of technology taking the human's role and in doing so disentangling the human-dog bond, as well as a dream scenario of technology augmenting human abilities arise. We argue that the current trajectory of digital technology for pets is increasingly focused on enabling remote interactions, an example of the nightmare scenario in our thematic analysis. It is important to redirect this trajectory to one of technology predominantly supporting us in becoming better and more informed caregivers.

*Index Terms*—human-animal interaction, multi-species households, digital ecologies, pet dogs, dog welfare

## I. Introduction

THE human-dog relationship is one of our[1] oldest known interspecies relationships (Shipman, 2021) and one could argue that technological advances have altered and improved the way we co-exist (Hill, 2018) from the start of our shared history. Only more recently have we started designing technology "for" dogs while arguably involving them in the design process (Aspling, 2020). Studies of the human-dog relationship have built a rich understanding of how we co-exist (cf. Mullin, 1999; Walsh, 2009; Gray & Young; 2011; Shir-Vertesh, 2012; Blouin, 2013; Gee & Mueller, 2019; Westgarth et al., 2019; Turnbull, 2020; Hawkins et al., 2021). Over the past decade, research attention into digital technologies that inform and complement human-dog relationships has similarly intensified, exploring topics from communication technologies (Paldanius et al., 2011; Lemasson, Pesty, & Duhaut, 2013), remote monitoring (Golbeck & Neustaedter, 2012), quantification of animal activity data Paasovaara et al., 2011; Weiss et al., 2013), to technology-mediated dog to dog communication (Hirskyj-Douglas et al., 2019; 2021). A major point in this increase of academic focus on technology in the animal domain has been the emergence of the field of *Animal-Computer Interaction (ACI)* and its guiding principle of 'animal-centered' design of technology (Mancini, 2011) encapsulated by the ethos of only designing technology that animals 'want or need' (Zamansky et al., 2017). Since its manifesto, a varied body of work has arisen designing technology for dogs, covering tangible technologies dogs physically interact with, as well as wearable, olfactory, screen-based, and tracking technologies (Hirskyj-Douglas et al., 2018).

Industry has caught up with, if not overtaken, the momentum of designing these technologies, with commercially built and marketed digital technologies for pets becoming more and more widespread, in particular smart feeders and toys and pet wearables (Wadhwani & Gankar, 2020). Research has shown that consumers increasingly spend to support their dogs (Dotson and Hyatt, 2008) and in the context of pet wearables are partially driven by ensuring their pets' physical safety (van der Linden et al., 2020). Commercial pet wearables, in particular, have attracted more research attention, covering aspects like the data they capture (van der Linden et al., 2019), their use in domestic (Väätäjä et al., 2018; Zamansky et al., 2019; Jayawardene et al., 2021) and professional settings (Zamansky & van der Linden, 2018), and what consumers find important in them (Ramokapane et al., 2019).

To better understand whether technology could, and would, support us in our ethical duties towards our companion animals (Yeates and Savulescu, 2017), or deal with the challenge of being 'responsible owners' (Westgarth et al., 2019) we need to take a critical eye towards the fundamental role of technology in the human-dog relationship. Lawson et al. (2015; 2016) were among the first to do so through design fiction based on the then state-of-the-art, identifying consumers having a strong desire for the use of technology, as well as urging several words of caution. Technological progress is unrelenting, and many of the once hypothetical devices are now marketed to dog owners around the world and increasingly appearing in their homes.

[1] Throughout this manuscript we will refer to "our" dogs to acknowledge the close ties and moral responsibility we have as human beings towards a species we domesticated and integrated into human society.

In its most recent report of 2021, the People's Dispensary for Sick Animals, a registered charity focused on pet welfare in the UK, found that there are 9.6 million pet dogs living in the UK, meaning that 26% of UK adults own at least one dog (PDSA, 2021).

It is estimated that the annual spending on pets and related products in the UK alone has increased significantly over the past years, now exceeding £9.5 billion (Office for National Statistics, 2022, p. 77). Beyond crude statistics and marketing reports, research has also increasingly found that spending on pet dogs increases as they become more integral parts of their adoptive human families (Brockman, 2001; Holbrook, 2001; 2008).

Research has repeatedly shown there is potential, and interest, for pet-dog technology beyond simple replacement of unwanted tasks (Paldanius et al., 2011; Hall et al., 2018), giving a clear reason as to why pet technology is similarly on a growing spending curve (Wadhwani & Gankar, 2020).

Yet, as dogs are deeply entangled with humans, their lives are frequently narrated from an anthropocentric perspective (Germonpré et al., 2018), and we should not disregard how this has set in motion a long history of control and power over dogs that has led to their commodification (Uerpmann & Uerpmann, 2017). Modern society has taken this to the extreme as "the capitalist commodification of animals is extensive" (Uerpmann & Uerpmann, 2017), including ownership of pet dogs valued as toys, status markers or for their 'brand' (Beverland et al., 2008). Modern social media further contributes both to the commodification of dogs as well as pushing capitalist narratives for goods and services dog owners need (Maddox, 2021).

We thus need to reconsider carefully the triadic human-dog-technology interactions (van der Linden et al., 2019; van der Linden, 2021) that may occur and the socio-cultural context in which those daily human-dog routines are framed (Hirsch-Matsioulas & Zamansky, 2020).

As Haraway said when interviewed regarding the fallacy of thinking about technology in absolute terms: "something is really seriously wrong and yet that's not all that's happening" (Gane, 2006, p. 151). Technophobia is a common attitude as new technologies keep emerging (Davidson et al., 2019), but not particularly constructive. Rather, we need careful consideration of how technology integrates into our lives and its potential harms, or lack thereof, to formulate any informed direction (Orben, 2020). This is especially important as the lines between digital and 'real' worlds becomes increasingly blurred and technology itself becomes not just a mediator between these realms, but a mediator of change in the real world per sé. Such blurring leads to additional challenges of understanding how human, animal, and technology intersect and affect each other in different cultural and socio-economic contexts (see e.g., Searle et al., 2021;), let alone how they give rise to fundamentally new 'digital animals' (Adams, 2020). And perhaps most challenging of all, how the asymmetry between human and animal in terms of the power and control they have over technology can be dealt with (see Kamphof, 2013). The purpose of this study is thus to carefully consider the role that technology is to play in the human-dog relationship without *a priori* technophobia or technophilia. We investigate the attitudes towards using technology across different activities that make up daily interspecies routines (e.g., feeding, walking, training) and the considerations that people make regarding its use.

## II. METHODS

### A. Participants

As there are many cultural differences in human-pet dynamics (Gray and Young, 2011; Hurn, 2012), we specifically chose to study a sample focused on one country to ensure in-depth understanding of the same surrounding factors (e.g., devices marketed to people, typical range of affordances relevant to dog ownership, and similar restrictions to daily routines from Covid-19 restrictions, covid restrictions). We recruited 155 participants using a purposeful sampling method, recruited via Prolific academic, where the participant information sheet was made available. All participants volunteered, and were reimbursed £0.50 for their participation in line with fair payment principles set out by Prolific at the time of the data collection. Inclusion criteria were (1) over 18 years of age, (2) a pet dog owner, and (3) British native English speakers.

### B. Procedure and online survey

Data were collected in March 2021. We received ethical approval (ref. 29162) from Northumbria University. The online questionnaire provided a full participant information sheet detailing the study, and required participants to explicitly give consent before proceeding.

The survey (see Appendix I) asked demographic questions regarding participants' and their dogs, and had participants describe their daily routine with their dog and their initial thoughts about using technology with them. Next, we described five different activities dog owners would engage in: (1) feeding their dog, (2) walking their dog, (3) playing with their dog, (4) cleaning up after their dog, and (5) training their dog. For each of these activities we gave a description of the activity and asked Likert scale questions whether they would be likely to use technology in that activity, how comfortable they would feel having that technology automate it, and open-ended questions (max 300 words) asking to explain their underlying rationales in detail. We purposefully did not overspecify what or how participants should interpret "technology" as (i.e., by restricting it to 'digital' or 'data-driven' technology), to ensure the data collected would be as reflective of what people cared for as possible. Finally, we used the Companion Animal Bond (CAB) scale (Poresky et al., 1987) to capture a measure of the strength of the bond between participants and their dog.

### C. Analysis

Quantitative data elicited were analyzed descriptively to provide an overview of the data. Qualitative data were analyzed using Braun and Clarke's (2017) Thematic Analysis protocol. These data encompassed a number of general questions including (1) people's routines with their dogs, (2) their initial thoughts where they might use technology, and (3) their specific rationales why they might (not) use technology across the five investigated activities. We took an inductive approach to generating codes for these three aspects looking at them primarily through the lens of anthrozoology (i.e., the study of how human and non-human animal relationships) and cyberpsychology (i.e., the study of how technology affects human behavior) to inform our thinking of both human-animal and human-technology relationships into a so far fairly unexplored human-animal-technology relationship perspective. All authors individually read through and became acquainted with the data, generating codes with initial definitions. We met three times to discuss and refine these codes, each time refining definitions, clarity, and scope, which led to integration, discarding, or creation of new codes based on theoretical insights.

## III. RESULTS

### A. Descriptive analysis

Human participants were an average age of 36 years of age (σ=14, youngest 18, oldest 69). 67% were female, 32% male, 1% undeclared. All were UK nationals, spoke English as their first language and had pet dog(s).

Nearly all participants had had their dogs for over a year (90%), with the remaining participants either having had them 6 to 12 months (9%) or 3 to 6 months (1%). Age-wise, there was a fair spread, with 6% of participants giving care to a juvenile dog, 27% to an adolescent dog, 43% to a mature dog, and 25% to a senior dog. Type of dog was similarly fairly spread, with only 2% of participants giving care to an extra small dog, 35% to a small dog, 34% to a medium sized dog, 27% to a large dog, and 2% to an extra large dog.

On average, the bond between participants and their dogs as indicated by the Companion Animal Bond (CAB) instrument was strong (M=4.0, σ=0.57, N=155).

As Figure 1 shows, likelihood of technology being used and comfort with automating were highly similar. Two of the five activities led to a polarized response: participants were generally generally likely (and trustful) to use technology for cleaning up after their dog (M=4, σ=1.3), while unlikely (and distrustful) to use technology for walking their dog (M=1, σ=1.1). The remaining three activities, feeding, playing with, and training dogs were ambivalent (M=3, σ=1.4, 1.3, 1.4) explained by a plurality of underlying rationales.

The label participants attributed to their relationship showed a split between those labelled themselves as caregiver (14%) (i.e., perhaps indicating a moral or virtue ethical approach towards animal stewardship), owners (47%) (i.e., perhaps indicating a typical kind of 'ownership' accompanied with 'property' views of the animal as is deeply ingrained in UK society (Srinivasan, 2011)), and anthropomorphized (35%) (i.e., perhaps indicating a relationship where the animal is incorporated into the human familial sphere (Shir-Vertesh, 2012). However, given how ingrained particular terms are to refer to human-dog relationships in the English language (i.e., 'owner' being a neutral term in common language) this should be taken with a grain of salt.

### B. Thematic analysis

Through the qualitative analysis of the data on routine and the wants and rationale for technology, we constructed a codebook with 19 codes for people's daily routines, 13 codes for people's reflections where they might use technology, and 24 codes for people's rationales why they might (not) use technology. Within the rationale codes, 11 codes related to technology resistance, 10 codes regarded technology receptiveness, and the remaining 3 codes were about technology ambivalence. Following finalization of the codebook, the first author re-coded all raw data.

Following further collaborative analysis sessions, these codes were condensed into five core themes. The focus here was on unique findings related to how technology slotted into the human-animal relationship. The final five themes, agreed upon by all authors, build on the individual codes and explore how habitual entanglement as a result of the human-dog dyadic relationship plays an important role in our conceptualization of, and expectations towards, technologies we consider for our dogs.

*Theme 1: Human and dog routines are deeply entangled. ("Entanglement")*

Our daily routines with pet dogs consist of a variety of recurring activities that lead to a strong habitual entanglement of our lives. From eating, napping and relaxing together, going for walks, or simply spending time together, many participants describe such activities as being undertaken together, painting a picture of a deeply intertwined routine, where dogs are very much considered as a fundamental member of the household: "*we watch tv together as a family with him and then he goes to bed*," **(P022).** Often their dogs were perceived as agents that actively choose if, when, and in which ways they want to entangle with us, too: "*she likes to 'help' in the garden by digging up holes while I'm planting veg /.../*" **(P078)**. Both prior examples provide an insight into the anthropomorphization some pet owners place upon their pets. Further, dogs are often actively drawn into human habits and routines such as eating at certain times ("*we go for another walk before tea, then he has his tea* [tea as a term is often used in lieu of 'dinner' or an evening meal in British English]", **P048**) or entertainment ("*he likes listening to classical music with me*", **P086**).

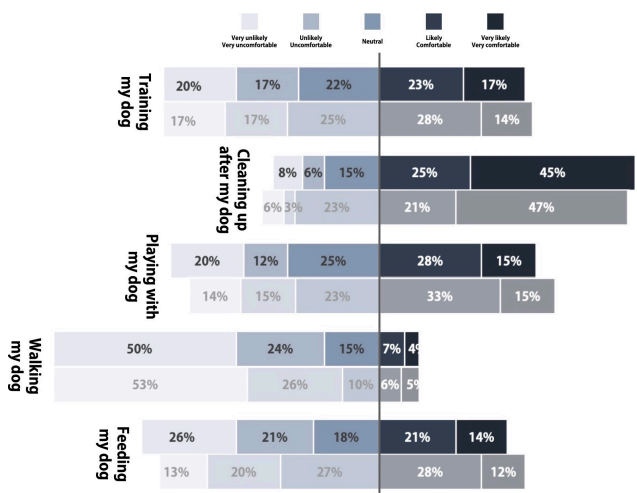

**Fig. 1**. Participants' attitudes towards technology in activities. Top bar of each activity indicates likelihood to use technology for that activity ("very unlikely" to "very likely"), bottom bar comfort with technology automating that activity ("very uncomfortable" to "very comfortable"). Results are aligned along agreement to visually indicate different attitudes among the activities.

It is not only our habits that entangle us, but our physical and temporal context and the social affordances it brings, both in private and public space, further entangle our lives. Our study focused on UK citizens, which did highlight a few fairly British cultural examples. For example, the affordance of many living in houses with private gardens or having green spaces nearby in more urban and city areas allowing convenience for dogs to dictate toileting and exercising times ("*they wake me up usually to go for a wee in the garden /.../ then go for a proper walk*", **P064**).

However, these routines naturally adapt as required, for instance, during the COVID-19 pandemic as participants spent increasing amount of time at home thus causing changes to our outdoor activities, where for months in the first lockdown we were not permitted to be outside for more than 30 minutes at a time (e.g., "*Pre COVID we would drive for a long 2-3 hour walk in the hills a few times a week but at [the] moment it's a hour long walk in surrounding fields*", **P151**). This of course impacted indoor activities, too (e.g., "*I've been at home through both lockdowns and he is back in the habit of sitting with me all day*", **P088**).

The notion of human and non-human lives becoming entangled is well established in anthropology (Mullin, 1999; Hamilton & Taylor, 2012; DeMello, 2012; Hurn, 2012) and multi-species ethnography (Kirksey & Helmreich, 2010; Pacini-Ketchabaw, Taylor, & Blaise, 2016). Indeed, "human nature is an interspecies relationship" (Tsing, 2012, p. 144).

We use entanglement here not just to mean co-habitation or co-existence, but to indicate that a human-dog dyad can increasingly be seen as an inseparable 'thing'. As Barad (2007) writes: "to be entangled is not simply to be entwined with another, as in the joining of separate entities, but to lack an independent, self-contained existence." In line with research which has argued our pets become parts of our extended selves (Jyrinki, 2005; Jyrinki & Leipamaa-Leskinen, 2005; van der Linden et al., 2019) we see such entanglement exactly as the human-dog relation being the fundamental 'thing' that exists.

Effectively, our dogs equally decide over our lives through the expectations they set for us, like when to go for walks (Holmberg, 2019), and even where to go for those walks (Dunnett, Swanwick, & Woolley, 2002, p. 43).

While there are certainly nuances to be taken in terms of complex power relations in human-animal relationships that are often in favor of humans (Giraud, 2021), we do not ignore such power relationships between the human 'owner' and their dog (Tuan, 1984), nor fail to understand that human and dog have different interests (Blouin, 2012). We use entanglement exactly because it reflects participants' answers about their "home culture" and how habits and daily ceremonies are shaped together with their dog.

Technology, as we will see later, becomes important to understand as assimilating technologies into this entanglement brings yet another potential source of agency affecting both human and dog: "the flow of agency moves both ways - we alter our tools as our tools alter us; we domesticate one another" (Hayler, n.d.).

### Theme 2. Entanglement builds close multi-species households. ("Closeness")

Whereas some people might only 'feed' their dogs, many participants noted actively 'eating' together ("*he then sits and eats lunch and dinner with me*", **P106**), and similarly, other activities become actively animal-initiated with dogs increasingly (perceived as) expressing their agency and interest, from obtaining food ("*in the evening he asks for treats when he comes into the living room,*" **P153**) to engaging in play ("*and play with her whenever she asks*", **P027**).

We found this clearly in our study with many participants talking about the amount of physical affection they gave to their dogs throughout the day ("*play and cuddle with my dogs whenever I have a spare moment*", **P007**), to showing affection deemed appropriate to them in other ways ("*talking to her and tickling her ears (she is very old)*", **P143**). Equally so, dogs show affection and emotional closeness or trust towards humans through seemingly simple acts as being comfortable enough to snooze around humans working from home ("*he then sleeps under my desk while I work*", **P075**).

Animals in multispecies households bring out affectionate behavior in humans (Walsh, 2009). Entanglement may initially be observed only as habitually and physically co-located events with human and animal involved, but they allow for multispecies households to emerge in their own right where pets actively participate in "doing" and being family (Irvine & Cilia, 2017). At the same time, dogs' body and disciplined performance mirror cultural constructions of 'home' and 'homemaking' (Power, 2012). Dogs, in our intimate relationships, very much decide over our life by setting out requirements for when we do things (Holmberg, 2019).

These seemingly mundane joint activities—especially when not simply co-located in time and space, but are perceived as emotionally joint activities create intimate human-dog bonds as we share dissonant but overlapping rhythms (Holmberg, 2019). This closeness becomes important to consider, as it brings with it a host of additional expected behavior and considerations in terms of our duty of care and the role that technology may play in it.

### Theme 3: Our entanglement and closeness lead to a priori conceptualizations of technology as things that ought to support us in our duty of care. ("Support")

When participants were asked to consider what kind of activities in their daily routines they might want to use technology for, a number of different potential roles came up, primarily in supporting our caregiving and easing our own worries stemming from our entangled lives and closeness as well as the responsibility humans took on themselves, personally as well as legally, once they became 'owners'.

When it comes to easing our worries, technology is perceived as something that could help us keep track of our pets when we cannot be there for them ("*I would like a camera so I can check in on her when I am out of the house /.../*"), to still ensure they are safe ("*/.../ and an app that links to [the] door would be good so she can go in the garden when I let her and I know she is back inside safe*", **P007**). Although occasionally, those with experience with such technology did express concern about its unsettling effect from the dog's point of view: "*I have a dog camera where you can communicate with your dog /.../* although when I did use it he was confused and a tad disconcerted as he didn't know where I was!" (**P140**). This may very likely be because our relationship with our dogs is such a physical, embodied one, that the sudden shift to a different, disembodied interaction is too much.

Indeed, technology should support us, and complement our caregiving abilities, but not negatively affect our emotions by, for example, compelling us to face our physical absence: "*I do not think I would like a visual device to monitor her whilst I am at work, as this would make me sad as I do not like leaving her.*" (**P045**). Beyond the immediate easing of worries, technology was more typically primarily conceptualized of as something that should help with our caregiving and understanding of our pets, either through providing us with quantified other (Nelson & Shih, 2017) information : "*we would love to try technology which measures her fitness levels as this is something we would like to improve upon,*" (**P069**), or by supporting us in the very interspecies activities we engage in and our own limitations therein.

Such limitations may be temporal, such as in the case of multispecies households where some participants noted the difficulty of juggling caregiving roles and responsibilities: "*twice a week I look after my grandchildren in their own home and often return after her dinner time so a device would be a great help to me but as she's a GSD it would have to be sturdy!*" (**P042**), or to do with our own limitations in providing our pets with the stimulation they deserve: "*[i would like] a laser projector that doesn't require human operation, since [my dog] loves chasing the laser. I don't mind playing with him with the laser, but he would be able to play with it more often if it didn't always require human operation.*" (**P092**). In effect, technology is *a priori* conceptualized as something that should maintain and promote our entanglement, and thereby, closeness, bringing with it the potential for strong reactions depending on whether it is seen to (dis)entangle.

***Theme 4: Reflecting on the role of technology in our entangled co-existence leads to nightmares of technology's potential disentangling effect. ("Nightmares")***

While only one activity, dog walking, received mostly negative views towards the use of technology, *all* activities had considerations on the negative impact using such technology could have. This seemed to occur, regardless of activity, primarily when the use or deployment of technology was seen as something that threatened our entanglement—specifically our bond by either replacing us, or reducing our involvement and responsibility and thereby slowly disentangling our routines. Some participants would clearly consider *all* technology to be unwanted for this reason, because "*if you are going to use technology in some way, then you are giving up quality time an owner usually spends with their pet,*" (**P096**) or even because "*if there is technology doing the jobs you should be doing as an owner, it means you may take less responsibility for the dog.*" (**P132**).

For most people, though, the reasons why technology is thought of as disentangling us varies on the activity and the context, having to do with factors related to more specific fears of (1) technology not being able to handle things, (2) technology replacing us in our role as caregivers, and (3) losing out on benefits gained through activities (e.g., improved fitness). Several activities are seen as being too complex to bother with technology. When it comes to feeding, issues with portion management, or type of feed are considered as impossible for technology to deal with: "*my dog requires a very specific diet of dry and wet food that I don't believe the dog food dispenser could give out*" (**P107**). In other activities, it is the dog's inherent agency and capacity to decide what and when to do that complicates matters, as for example, one participant noting that "*my dog usually goes straight outside after eating to go to the toilet. If we were out there wouldn't be anybody here to let her out.*" (**P096**) When it comes to training our dogs, participants note that it is too complex of an affair to leave it up to technology, and that "*human teaching is always going to be more beneficial, as we can adapt to the dogs behaviour and what it needs, rather than following a designated protocol. Using a device to do so, would work for some instances but not all, and it would probably cost a lot of money.*" (**P081**)

Among those people who accepted that technology *could* work, other concerns became apparent across different activities again. Rather than fearing that a smart feeder would break down or not handle feeding well, some were concerned about losing the interspecies interaction per se: "*my dog gets excited for her breakfast and dinner and I love how she gets all happy around the same time each day and sits where her food is kept. I love how pure that is and I wouldn't wanna lose that to some technology on my phone.*" (**P047**) Training, in particular, is seen as an activity that shapes and strengthens the bond ("*i would rather train my dog myself, as it becomes more comfortable with the people who look after it during training*", **P015**) and something where even if technology "*sounds helpful and time saving, but would my dog then respect and listen to me or the device? I would be cynical about this.*" (**P056**)

Finally, the last fear participants raised across several activities was technology taking away benefits gained through doing these activities. Even though most dog owners would gladly let technology clean up after them, many also wondered "*It is not the nicest job [cleaning up after our dog], but it can be a big indication something is wrong. I would worry about relying on technology too much, I would miss a sign it was maybe runny or a different colour, or not as frequent.*" (**P017**). There are also benefits to ourselves we fear we might miss out on, as many participants noted that dog walking has benefits for their own fitness, so much that "*the main reason I have a dog is to get outside and get me exercising*" (**P010**) was a recurring sentiment.

Technology, effectively, is conceptualized as a negative force with the potential to lead to a nightmare scenario where it disentangles us and takes away the benefits that our multispecies households give us—the companionship, entertainment, and protection that provides us with a mutual multi-faceted security in our own mental and physical lives (cf. Turnbull, 2020 ). While 'becoming with' (Haraway, 2013) the dogs, owners learn the needs and desires of the dogs and the dog's repetitive behaviors throughout the day (Corkran, 2015). In this sense, owners perceive themselves as a knowledge authority about the dog, one that technology cannot replace. Given that there have been cases of pets going hungry when smart feeder infrastructure broke down (BBC, 2020), fears for critical technology breaking down might indeed too be well grounded.

***Theme 5: Reflecting on our role in the multispecies household leads to dreams of technology's potential to increase our ability to provide care. ("Dreams")***

Views on technology are not always negative. Across different activities, positive conceptualizations on technology's role in our relationship occurred, having to do with its potential to support us in our caregiving and deal with our own limitations. For example, when it comes to walking our dogs, for some technology "*would be a godsend as sometimes my illness limits my walking*" (**P014**), while others similarly realized that "*[my dog] would have (in her youth) enjoyed more walks than what any human could have given her.*" (**P143**) Other physical activities like play are similarly seen as something where we feel that we could do better for our dogs: "*I can't always give him the amount of play that he needs, so any sort of play device is great*" (**P014**) who could always use more ("*I am capable of playing with my own dog, but he would never turn down more play time!*", **P016**), and establish affordances taken away by our need to e.g., work, as "*my dog sometimes wants to play when I'm trying to work, so this may come in handy*" (**P029**) Technology is similarly conceptualized as something that could help us ensure consistency in caregiving, like ensuring that "*he is getting food at the exact same time every day*" (**P075**), especially if we cannot guarantee doing so: "*I am terrible at remembering to feed my dog at set times. So I would like a device that does it automatically.*" (**P023**)

Most positive conceptualizations of technology seem to center on what the above reveals: giving us support where we need it, whether that need arises due to physical disabilities, lack of time, or knowledge. Technology, quite understandably from the fears we identified, should reinforce our entanglement and increase our ability to provide care. And perhaps, indeed, whether we would use technology for a particular technology is really related to what that activity is to us and them. For example, when it comes to cleaning up after them, "*your dog doesn't care who does it.*" (**P143**). Even with technology enabling good passages from disability to ability, however, we see that there is a limit to how much we are willing to rely on technology to address our own in- or disabilities before we consider the technology to have replaced us as the dog's caregiver, and in doing so disentangled our relationship.

IV. GENERAL DISCUSSION AND CONCLUSION

Our thematic analysis led us to construct a theoretical understanding of the role of technology in human-dog relationships. As Figure 2 shows more clearly, human and dog routines are strongly entangled and there is a desire for technology of some kind in these relationships. Yet, while technology is typically thought of as something that should help support our caregiving, the technology available on the market seems to increasingly be focused on replacing, rather than supporting us.

*A. The root of the issue with technology's role in the human-dog relationship*

Lawson's et al. (2015; 2016) words of caution against the uncritical design and proliferation of technology in human-dog relationships were in large part spurred by having observed a then strong desire for technology among dog owners which they argued had the potential to "*undermine human-animal bonds.*" In particular, they warned against assuming "relatively simple" technology can usefully enhance our co-evolved innate ability to interpret cross-species behavior and psychological states. Technology, they warned, should not substitute for human interpretive aspects of human-dog relationships. However, this may be an oversimplification, as while some evidence shows untrained people can identify dog emotions (Bloom and Friedman, 2013), when it comes to identifying behavior, those without theoretical knowledge frequently perform poorer in identifying behavior or behavioral cues (Tami and Gallagher, 2009; Bloom and Friedman, 2013; Demirbas et al., 2016). We should thus not discount technology that supports people in understanding their dog's behavior.

Our findings extend Lawson et al.'s understanding in two important ways. First, we found no clear evidence for people still having a strong desire for pet technology—even though more advanced technology has come on the market. While technology to support in chores, like cleaning up after them, is welcomed, dog owners' attitude to using technology in most other activities (e.g., training, playing) is ambivalent or even negative, with technology primarily conceptualized as augmenting our caregiving capabilities, that is, adding to our own interpretation, rather than substituting it. As the market of pet technologies is growing (Wadhwani & Gankar, 2020), this difference from Lawson's earlier observations may be related to these technologies becoming more mainstream and the initial hype cycle around it having faded (Linden & Fenn, 2003).

Second, through our thematic analysis, we built a clearer understanding from our participants' perceptions *how and why* human-dog bonds may be undermined through technology. Our findings show that dog owners conceptualize technology as *augmenting* their caregiving capabilities, rather than fulfilling them in our place. This echoes insights from research on technology augmenting our own abilities beyond what we can physically do (whether due to physical, environmental, or temporary disability) and in doing so (re)gain autonomy and a certain extent of control and responsibility (Moser & Law, 1999). These traits (our abilities, autonomy, control, and responsibility) are part of the traits that we perceive as being owners of dogs: our duty of care to our dogs means we are supposed to do things with them and for them, to decide for them, and to control them. Technology, then, can be perceived as our own material and symbolic extension and to increase our own abilities.

However, this augmentation can only work if passage from disability to ability is smooth, or a "good passage" from disability to ability (ibid, 1999). If the technology we rely on fails, for example, by not being tuned to the dog's needs, infrastructural failures (see BBC, 2020) or other technological and material idiosyncrasies, it will not only *not* extend our own ability, it will create new disabilities that challenge some of the socially constructed characteristics of western dog ownership from Perry, Scott, & McKinley's (1997) work on accessibility, we conceptualize disability in the context of dog-human dyads as an impairment in our ability to pick up relevant sensory information (e.g., noticing our dog's behavioral cues), perform some action involving movement (e.g., throwing a ball, fill a feed bowl), or performing some cognitive function (e.g., understanding our dog's behavioral cues). Technology can provide a passage from such *dis*abilities to ability by augmenting our physical and mental abilities.

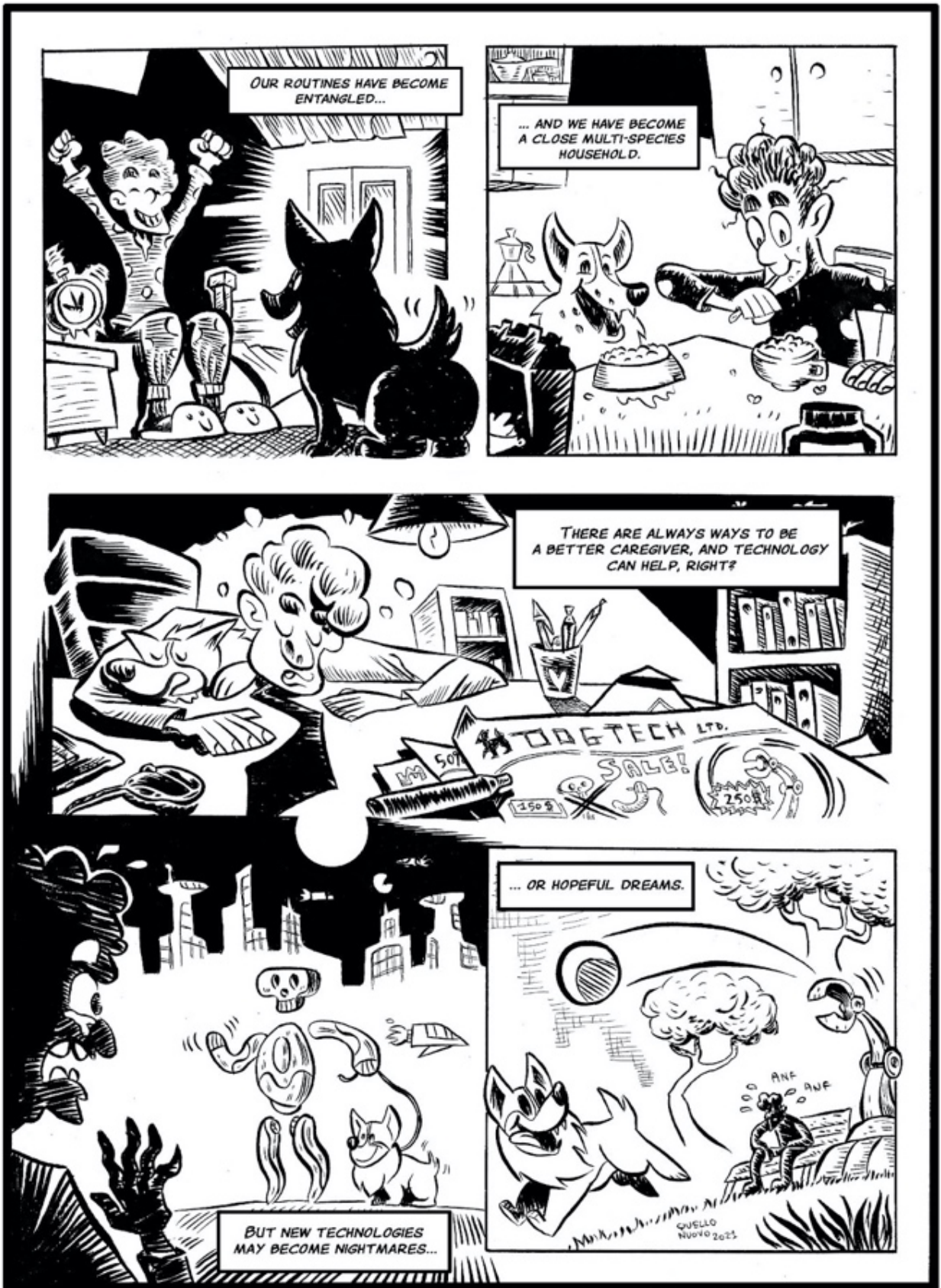

**Fig. 2.** Synthesis of the thematic analysis' most salient insights: entangled multispecies households and a priori assumptions of technology being for caregiving lead to parallel nightmares of being replaced by technology, and dreams of being supported by technology.

An important distinction to make is that of our *in*abilities in the dog-human relationship. Inabilities are anything a person *cannot* do (ibid, 1997). In the context of human-dog relationships, while there may be *in*abilities rooted in our own physical or mental disabilities, by and far the most common *in*ability is simply that we are not capable of being there. For example, because we have to work, or travel, or are otherwise occupied. Technology can, and does, provide some passage from this *in*ability to ability, but in doing so does not meaningfully augment the human's capabilities in the dog-human dyad but rather temporarily replaces them, thereby fulfilling the nightmare of disentanglement.

Bad passages, thus, we would argue, are much more likely to happen when technology is designed from the start not intending to bridge a gap from *dis*ability to ability (e.g., help us provide a better diet, play for longer), but to move from *in*ability rooted in us not being there to ability directly (e.g., feed our dog through a device when we leave them at home alone while at work or even on vacation). This recalls Weizenbaum's (1976) apprehension of our growing reliance on technology in lieu of social innovation. Consider his later reflections on his own work introducing technology to "improve" banking efficiency: "*What the coming of the computer did, 'just in time,' was to make it unnecessary to create social inventions, to change the system in any way. So in that sense, the computer has acted as fundamentally a conservative force, a force which kept power or even solidified power where it already existed.*" (Ben-Aaron, 1985)

Some Animal-Computer Interaction research has argued along similar lines, noting that social change in the domestic sphere could be argued to be more beneficial to the animal, as "*in an animal-centered approach, this problem might be solved such that dogs are never socially isolated, rather than the development of a technological solution to employ while they are left alone*" (Grillaert & Camenzind, 2016). Yet, this has had little effect on the trajectory of technology designed since then.

While in this research we specifically chose to investigate what role technology should play in human-dog relationships in general to provide a wider picture that may positively steer where technology goes, further empirical research exploring the lived experiences with concrete technologies is equally as necessary.

Qualitative, in situ research exploring the digital ecologies of human, dog, and technology and the concrete experiences therein open up new approaches to understand and explore how human-dog relationships change in the 'digital age'. Such studies would not only deepen our understanding of the notions of how technology may mediate agency and power in human-dog relationships, but are also more likely to avoid being affected by general attitudes of technophobia or technophillia.

As researchers, we can also consider the longer-term effects of these technologies on the human-dog relationship by monitoring how technology affects our behavior (e.g., do pet wearables help get pets and humans fitter?; do buttons sold to ostensibly allow dogs to 'communicate' their needs to humans (Judkis, 2021) increase interspecies understanding?). Similarly, it is critical to consider these technologies in wider frames of reference, such as security and privacy—are these technologies safe and secure? How invasive are any apps? How are data stored and used, and what are the ramifications should these get lost?

Such research leads into considering to what extent novel digital ecologies give rise to nightmare or dream scenarios, or just as possible, how humans adapt and use these technologies beyond their intended use to support them in achieving dream scenarios. This naturally would lead to new theories and approaches for the integration of digital technologies into multi-species households, while ideally ensuring research on their design and deployment are conducted ethically and safely.

*B. The problematic trajectory of technology for the human-dog relationship*

The current trajectory of technology for the human-dog relationship does little to address the root of this issue. While market research indicates that the trajectory of technology for dogs is driven by an increased spending on digital technology centered around health and wellbeing (Wadhwani & Gankar, 2020), in practice it results in pet wearables, smart pet toys, and smart pet feeders and bowls growing the most—all of which essentially enable remote monitoring and remote engagement with dogs, rather than providing "on the scene" support. This is exactly the kind of technology that could be used in our stead, rather than support us. During the Covid-19 pandemic many people adopted dogs, leading to many new multi-species households (Morgan et al., 2020; Jezierski et al., 2021; Holland et al., 2021) where dogs diminished our feelings of isolation and loneliness (Bussolari et al., 2021). Yet, now with increasing attempts at a "return to normal", that is, returning to our places of work and leaving dogs at home alone, many might seek for technology that could minimize the emotional pain that typically results from our separation (Kanat-Maymon et al., 2020), regardless of its longer-term effect on the human-dog relationship, as we have also seen in emotion-driven choices where giving away valuable personal data of both human and dog is seen as a fair trade-off for a sense of reassurance and security in always knowing where one's dog is (van der Linden et al., 2020).

For example, new technologies occur that aid in training, promising that "when it's time for you to take them for a walk, you can focus on the good times" (Companion, 2021), which may sound beneficial given the benefits of dog walking (Campbell et al., 2016; Koohsari et al., 2021), but seem to conflict with our finding that many dog owners would fear this technology replacing them during vital bonding moments of training and undermine the relationship. Another recent development is a smart camera-enabled ball for remote play promising to allow you to "interact with your pets from anywhere" (PlayDate, 2021). Yet, again, our findings would show such technology hits uncomfortably close to the nightmare theme of using technology to replace our inability to play with our dogs, rather than complement our ability to play more *with* our dogs.

Academia similarly seems on a trajectory towards technology that may further disentangle the bond, whether automating the identification and rewarding of good behaviors (Stock & Cavey, 2021), or holding workshops to build technology that considers disembodied remote interspecies communication as "doing activities together" (Väätäjä et al., 2017), and in doing so, further normalize and entrench physical separation. Such ideas do not remain as design fictions, as *DogPhone*, a recently designed prototype for dogs to (allegedly) initiate video-calls to their owners (Hirsky-Douglas et al., 2021) shows. To reiterate, our findings clearly show that such technology would fall into nightmare scenarios of replacing us, increasing our own anxiety at not being there for our dogs, and normalizing not *having* to be there for them. Interestingly, in the extensive media coverage given to DogPhone's promise a tension between wanting to do right, and perhaps not having considered how technology plays a role in the human-dog relationship becomes apparent.

In an interview with The Guardian the DogPhone's creators justified its design and contrasted it from other technology on the market because "*All this [existing] technology allows you to measure your pets' steps or ring your pets or remotely give your dog food, but your dog doesn't really have any choices*" (Davis, 2021). We would disagree with the notion of dogs not having a choice, given the research arguing how dogs have power to some extent to dictate our routines (Holmberg, 2019), let alone operate in the human world and use its infrastructure in their own right (Lemon, 2015). Of course, such power is reliant on an empathic 'owner', who, in most human societies, legally holds all power and can subvert any need or want of the dog. Yet, the people who would use such technology are likely already such empathic 'owners' susceptible to dogs' expression of wants and needs. More importantly, though, in the same interview, the authors also admitted that "*the device had actually caused her [the author] some anxiety*" (Davis, 2021)—reinforcing the point that this technology was made, likely with the best of intentions to do good for the dog, but could have benefited from more upfront anthrozoological research into the role of technology in the human-dog relationship. Coverage of *DogPhone* in the New York Times (Chung, 2021) showcased the challenge of media in covering these technologies while having to oversimplify for space and audience, as while the behavioral experts they cited who questioned how such technology could realistically help with isolation and separation anxiety, little space or time was left to question the dearth of research on the role of technology in the human-dog relationship upfront and address the elephant in the room: *why do so many pets need to be left at home, and is telepresence technology really the best way to address that?*

Rather than accept its current trajectory spelled out by the lack of clear grounding in an understanding of human-dog relationships, and the use of technology to jump from *in*ability from not being there to ability, we should work to redirect it. Technology's role in the human-dog relationship should be to positively and constructively address the issues we face together. For just one example, it should attempt to tackle the epidemic of dog obesity (Courcier et al., 2010; German et al., 2018) and its likely relation to our own weight and diet (Linder et al., 2021). Not by taking us out of the equation and automating our dogs' diets and exercise through some technological artifact, but by making us more informed and attentive caregivers, by understanding the underlying challenges we have in coping with pet food insecurity (Arluke, 2021), motivating us to become more active together (Zamansky et al., 2019), and perhaps through sharing data it captures for clinical and academic research (Lee & Lee, 2015). It should support our ability to physically engage with our dogs as we both grow older. It should help us train them well. It should help us understand their behavior and needs. But none of this should require having to replace ourselves with technology as a matter of routine.

Rather than go with the current trajectory of technology that replaces us and allows us to avoid having to make hard decisions regarding our moral duty towards our dogs, we should strive to understand better how technology could lead to better functioning and more equitable human-dog relationships. To make it very clear: the bottom line of our argument here is not even driven by an underlying acceptance of abolitionism (Francione, 2018), that is, rejecting entirely *any* animal 'use' by humans. It is reflecting on what we owe to our dogs, accepting that indeed, "*the planet is lousy with pets*," (McWilliams, 2016) and from there accepting that at the very least we ought to stop bringing dogs into domestic spheres where we *need* technology to provide adequate care for them. Is that too much to ask?

### ACKNOWLEDGMENT

The authors thank Andrea "Quello Nuovo" Menichini for the original comic artwork commissioned to describe our results far better than we could. Another major thank you to the reviewers of this manuscript who helped improve it significantly with their constructive suggestions. A final special thank you to all the owners who participated in this study.

**Dirk van der Linden** received the PhD from Radboud University in 2015. He is a Senior Lecturer with Northumbria University, UK. His research interests include requirements engineering and technology for animals.

**Brittany I. Davidson** received the PhD from the University of Bath in 2020. She is a Lecturer with the University of Bath. Her research interests include psychology and behavioral analytics.

**Orit Hirsch-Matsioulas** received the PhD from Ben-Gurion University in 2019. She is a Research Fellow with the University of Haifa, Haifa, Israel. Her research interests include anthropology and human-animal relations.

**Anna Zamansky** received the PhD from Tel Aviv University in 2009. She is an Associate Professor with the University of Haifa, Haifa, Israel. Her research interests include information systems and technology for animals.

APPENDIX I — SURVEY

*1. You and your dog*

We will first ask you a couple of questions to characterise your relationship with your dog.
- What size is your dog?
    - Extra small (like Chihuahuas)
    - Small (like Corgis or Dachshunds)
    - Medium (like Bulldogs or Poodles)
    - Large (like Boxers or Retrievers)
    - Extra Large (like Great Danes)
- How old is your dog?
    - Juvenile (Puppy)
    - Adolescent
    - Mature
    - Senior
- How long have you had your dog?
    - 0 to 3 months
    - 3 to 6 months
    - 6 to 12 months
    - More than a year
- How would you describe your relationship to your dog?
    - I am their caregiver
    - I am their owner
    - I am their parent
- Please give a brief description of your daily routine with your dog—what activities do you typically engage in together? (max 200 words)
    - Open text
- Are there any activities you would specifically like to use technology for, or would absolutely not want to? If so, why? (max 200 words)
    - Open text

*2. You, your dog, and technology*

An increasing number of dog-centred technologies are available to help you out in different daily activities with your dog. Below we will describe a couple of typical activities in which you may engage with your dog, and ask you about your thoughts of using technology for those activities.

(1) Feeding your dog: imagine technology that helps you refill your dog's feeding bowl through an app, or even refills it automatically.
- Assume you were given such a device for feeding your dog. How likely would you be to use it?
    - 5pt Likert scale anchored with "Very unlikely" and "Very likely"
- Assume you were using such a device. How comfortable would you be for it to automate feeding your dog?
    - 5pt Likert scale anchored with "Very uncomfortable" and "Very comfortable"
- Please briefly explain your underlying reason(s) for your answers here [max 200 words]
    - Open text

(2) Walking your dog: imagine technology that could help you walk your dog, or even does it automatically.
- Assume you were given such a device for walking your dog. How likely would you be to use it?
    - 5pt Likert scale anchored with "Very unlikely" and "Very likely"
- Assume you were using such a device. How comfortable would you be for it to automate walking your dog?
    - 5pt Likert scale anchored with "Very uncomfortable" and "Very comfortable"
- Please briefly explain your underlying reason(s) for your answers here [max 200 words]
    - Open text

(3) Playing with your dog: imagine technology helps you play with your dog, or even plays with it automatically.
- Assume you were given such a device for playing with your dog. How likely would you be to use it?
    - 5pt Likert scale anchored with "Very unlikely" and "Very likely"
- Assume you were using such a device. How comfortable would you be for it to automate playing with your dog?
    - 5pt Likert scale anchored with "Very uncomfortable" and "Very comfortable"
- Please briefly explain your underlying reason(s) for your answers here [max 200 words]
    - Open text

(4) Cleaning up after your dog: imagine technology that can help you clean up after your dog's mess, or even do it for you automatically.

- Assume you were given such a device for cleaning up after your dog. How likely would you be to use it?
  - 5pt Likert scale anchored with "Very unlikely" and "Very likely"
- Assume you were using such a device. How comfortable would you be for it to clean up after your dog?
  - 5pt Likert scale anchored with "Very uncomfortable" and "Very comfortable"
- Please briefly explain your underlying reason(s) for your answers here [max 200 words]
  - Open text

(5) Training your dog: imagine technology that helps you train your dog, or even do it for you automatically.
- Assume you were given such a device for training your dog. How likely would you be to use it?
  - 5pt Likert scale anchored with "Very unlikely" and "Very likely"
- Assume you were using such a device. How comfortable would you be for it to automate training your dog?
  - 5pt Likert scale anchored with "Very uncomfortable" and "Very comfortable"
- Please briefly explain your underlying reason(s) for your answers here [max 200 words]
  - Open text

3. *You and your dog, again*

Finally, we will ask you some questions to understand your relationship with your dog. For each of the below statements, answer how often you typically do those things.
- How often are you responsible for your dog?
  - 5pt Likert scale anchored with "Never" and "Always"
- How often do you clean up after your dog?
  - *ibid*
- How often do you hold, stroke, or pet your dog?
  - *ibid*
- How often does your dog sleep with you?
  - *ibid*
- How often do you feel that your dog was responsive to you?
  - *ibid*
- How often do you feel that you had a close relationship with your dog?
  - *ibid*
- How often do you travel with your dog?
  - *ibid*
- How often do you sleep near your dog?
  - *ibid*

4. *About you*
- To what extent would you say you use technology for activities in your day to day life?
  - 5pt Likert scale anchored with "Never" and "Always"
- To what extent would you say you generally feel comfortable letting technology automate things for you?
  - 5pt Likert scale anchored with "Very uncomfortable" and "Very comfortable"